\documentclass[twocolumn]{aastex631}

\newcommand{\lum}{erg~s\ensuremath{^{-1}}}
\usepackage{epsfig,graphics,subfigure,psfrag,amsmath,amssymb,amscd}

\begin{document}

\title{Dissonance in harmony: The UV/optical periodic outbursts of ASASSN-14ko exhibit repeated bumps and rebrightenings}

\author[0000-0001-7689-6382]{Shifeng Huang}
\affiliation{CAS Key Laboratory for Research in Galaxies and Cosmology, Department of Astronomy, University of Science and Technology of China, Hefei, 230026, China; sfhuang999@ustc.edu.cn, jnac@ustc.edu.cn}
\affiliation{School of Astronomy and Space Sciences,
University of Science and Technology of China, Hefei, 230026, China}

\author[0000-0002-7152-3621]{Ning Jiang}
\affiliation{CAS Key Laboratory for Research in Galaxies and Cosmology, Department of Astronomy, University of Science and Technology of China, Hefei, 230026, China; sfhuang999@ustc.edu.cn, jnac@ustc.edu.cn}
\affiliation{School of Astronomy and Space Sciences,
University of Science and Technology of China, Hefei, 230026, China}

\author[0000-0001-5012-2362]{Rong-Feng Shen}
\affiliation{School of Physics and Astronomy, Sun Yat-sen University, Zhuhai 519082, People's Republic of China; shenrf3@mail.sysu.edu.cn}
\affiliation{CSST Science Center for the Guangdong-Hong Kong-Macau Greater Bay Area, Sun Yat-sen University, Zhuhai 519082, People's Republic of China}

\author[0000-0002-1517-6792]{Tinggui Wang}
\affiliation{CAS Key Laboratory for Research in Galaxies and Cosmology, Department of Astronomy, University of Science and Technology of China, Hefei, 230026, China; sfhuang999@ustc.edu.cn, jnac@ustc.edu.cn}
\affiliation{School of Astronomy and Space Sciences,
University of Science and Technology of China, Hefei, 230026, China}

\author[0000-0001-6938-8670]{Zhenfeng~Sheng}
\affiliation{Institute of Deep Space Sciences, Deep Space Exploration Laboratory, Hefei, 230026, China}

\begin{abstract}

ASASSN-14ko was identified as an abnormal periodic nuclear transient with a potential decreasing period. Its outbursts in the optical and UV bands have displayed a consistent and smooth ``fast-rise and slow-decay" pattern since its discovery, which has recently experienced an unexpected alteration in the last two epochs, as revealed by our proposed high-cadence \emph{Swift} observations. The new light curve profiles show a bump during the rising stages and a rebrightening during the declining stages, making them much broader and symmetrical than the previous ones. In the last two epochs, there is no significant difference in the X-ray spectral slope compared to the previous one, and its overall luminosity is lower than those of the previous epochs. The energy released in the early bump and rebrightening phases ($\sim10^{50}$~erg) could be due to collision of the stripped stream from partial tidal disruption events with an expanded accretion disk. We also discussed other potential explanations, such as disk instability and star-disk collisions. Further high-cadence multiwavelength observations of subsequent cycles are encouraged to comprehend the unique periodic source with its new intriguing features.

\end{abstract}

\keywords{Tidal disruption (1696) --- Supermassive black holes (1663) --- Black hole physics (159) --- Accretion (14)}

\section{Introduction} \label{sec:intro}
 Quasiperiodicity is a phenomenon that is often observed in active galactic nuclei (AGNs). OJ 287 is a well-known example of this, exhibiting a 12 yr quasiperiodic oscillation (QPO) in its optical light curve \citep{Sillanpaa1988}. This QPO is thought to be caused by a supermassive black hole binary \citep[SMBHB;][]{Lehto1996,Valtonen2008}. Furthermore, PG 1553 + 113 \citep{Ackermann2015} and SDSS J143016.05+230344.4 \citep{Jiang2022} have both been observed to show QPOs, and are likely to host SMBHBs. However, some AGNs display short-term QPOs in their X-ray emission, such as RE J1034+396, which has a QPO of the order of 1 hr \citep{Gierlinski2008}. These short-term QPOs in X-rays may be generated by the inner disk \citep{Jin2021}, while long-term QPOs may be the result of SMBHBs \citep{Huang2021}.

Recently, another quasiperiodic phenomenon in X-rays known as quasiperiodic eruption (QPE) has been discovered in GSN 069 \citep{Shu2018,Miniutti2019}. This signal is only present in the low state and disappears in the high state \citep{Miniutti2023a,Miniutti2023b}. Five other sources with QPEs have been identified, such as RXJ1301.9+2747 \citep{Sun2013,Giustini2020}, eRO-QPE1, eRO-QPE2 \citep{Arcodia2021}, XMMSL1 J024916.6-041244 \citep{Chakraborty2021}, AT2019vcb \citep{Quintin2023} and Swift J023017.0+283603 \citep{Evans2023,Guolo2023}. Additionally, a smaller type of recurrent X-ray eruption sources, called ultraluminous X-ray bursts (UXBs), were discovered by \cite{Irwin2016}.

ASASSN-14ko is a nuclear transient located in the Seyfert 2 galaxy ESO 253–G003, which was discovered in 2014 \citep{Holoien2014}. It has been observed to have a strong periodicity in its UV/optical light curves, with a period of 115 days and a negative period derivative of -0.0026 \citep{Payne2021,Payne2023}. The light curves of each outburst have a ``fast rise and slow decay" pattern, and the optical spectra show a blue wing in H$\beta$ during the flare that is absent in the quiescent state \citep{Payne2021}. It is thought to be related to a partial tidal disruption event \citep[TDE;][]{Payne2021,Payne2022,Tucker2021}, which is caused by an evolved star on an eccentric orbit \citep{Liu2023}. Additionally, \cite{Cufari2022} suggest that the disrupted star originated from a binary system via the Hills mechanism. \cite{Metzger2022} propose that the periodic behavior may be due to the mass transfer in a coorbiting extreme mass ratio inspiral system consisting of a supermassive black hole (SMBH) and a binary. \cite{King2023} believes that mass transfer between the white dwarf and SMBH can induce such periodic variability, which was previously suggested by \cite{Shen2019} to explain UXBs.  However, it should be noted that ASASSN-14ko has a black hole mass of $10^{7.85}\, M_\odot$ \citep{Payne2021}, and such a black hole would swallow a white dwarf without disrupting it. Alternatively, \cite{Sukova2021} argue that the periodic outburst could also be produced by the periodic perturbation of a star in the accretion flow. Similarly, star-disk interaction is also thought to could produce this phenomenon \citep{Linial2023b}.  

Rebrightenings are a common phenomenon in TDEs when the sources are monitored for a long enough duration \citep{Jiang2019,Yao2023}. They can be caused by the circularization of debris and delayed accretion following the initiation disruption \citep{Chen2022,Wang2023}. Other potential explanations for rebrightenings include partial TDE \citep{Wevers2023,Malyali2023,Liu2023b} and SMBHB systems \citep{Liu2014,Shu2020}. It is worth noting that, in addition to the rebrightening observed after the primary peak, an early bump during the rising stage has also been detected in AT2020wey \citep{Charalampopoulos2023}. Partial TDEs offer us a great opportunity to observe the entire process across multiple wavelengths due to its repeatability. Consequently, we have requested several target of opportunity (ToO) observations from \emph{Swift} to conduct multiwavelength observations on ASASSN-14ko. In this paper, we present new observations of this source and newly discovered phenomena. In Section \ref{sec:data}, we describe the data reduction and the multiwavelength light curves. In Section \ref{sec:ananlysis}, we analyze the X-ray spectra and the multiwavelength light curves. Finally, we discuss the physical mechanism of early bumps and rebrightenings in Section \ref{sec:diss}. In this work we assume cosmological parameters of $H_0=70\,\text{km}\text{s}^{-1},\text{Mpc}^{-1}$, $\Omega_{\rm M}=0.3$, and $\Omega_{\Lambda}=0.7$ in this work.

\section{Data Reduction and Results} \label{sec:data}
\subsection{Swift X-Ray photometry}
We use the X-Ray Telescope (XRT) on board the Neil Gehrels \emph{Swift} Observatory to perform X-ray photometry of the object, with a sensitivity range of 0.3 -- 10.0 keV \citep{Burrows2005}. To monitor its multiwavelength behavior, we have requested 10 ToO observations with high cadence starting from 2022 December 2. We also retrieve the public data (PI: Payne) from the High Energy Astrophysics Science Archive Research Center (HEASARC) website. The \emph{Swift} data are processed with \texttt{HEASoft 6.31.1}. To obtain the products (e.g., light curves and spectra), we run the task \texttt{xrtpipeline}, and then \texttt{xrtproducts}. The source region is chosen as a circular region centered on the object with a radius of $47.1^{\prime\prime}$, and the background region is chosen as a nearby circular region with a radius of $120^{\prime\prime}$.

\subsection{Swift UV/Optical photometry}
The UV/Optical Telescope (UVOT) is another instrument in the \emph{ Swift} observatory, which is equipped with seven filters (\textsl{V}, \textsl{B}, \textsl{U}, \textsl{UVW1}, \textsl{UVM2}, \textsl{UVW2} and white) \citep{Roming2005}. We run \texttt{uvotimsum} to sum up the images and then execute the task \texttt{ uvotsource} to generate the light curves with the source and background regions defined by circles with radii of $10^{\prime\prime}$ and $40^{\prime\prime}$, respectively. Following \cite{Schlafly2011}, we derive the Galactic extinction value of $E(B-V)=0.043$ using an online tool\footnote{\url{https://irsa.ipac.caltech.edu/applications/DUST/}}, and then we corrected the magnitude for each band using the extinction law of \cite{Cardelli1989}. In this work, we adopt the Galactic extinction values of 0.14, 0.18, 0.21, 0.27, 0.42, and 0.41 mag for the \textsl{V}, \textsl{B}, \textsl{U}, \textsl{UVW1}, \textsl{UVM2}, and \textsl{UVW2} bands, respectively.

\begin{figure*}[htb]
  \centering
  \subfigure[]{
  \includegraphics[width=0.7\textwidth]{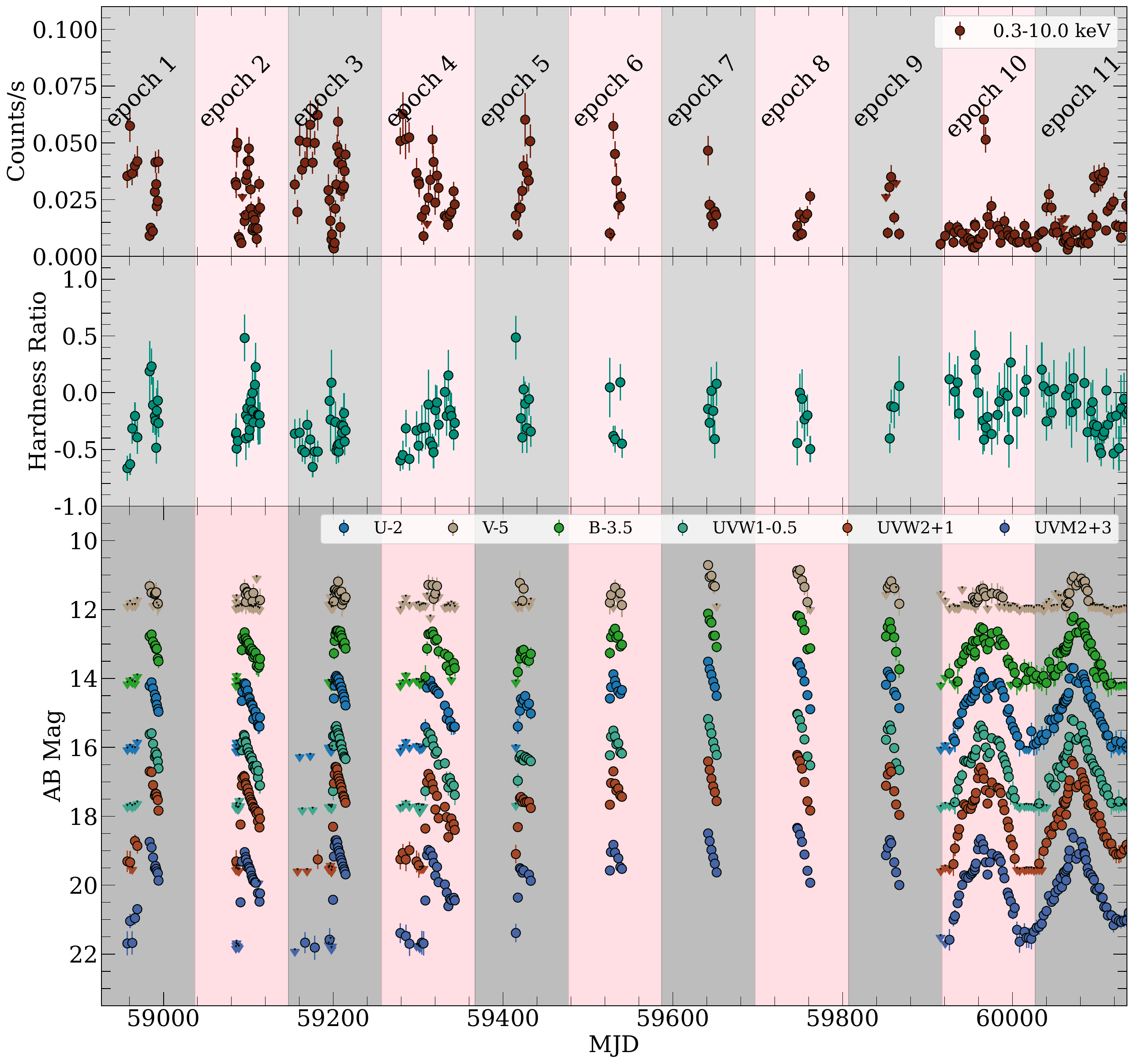}}
  \subfigure[]{\includegraphics[width=0.85\textwidth]{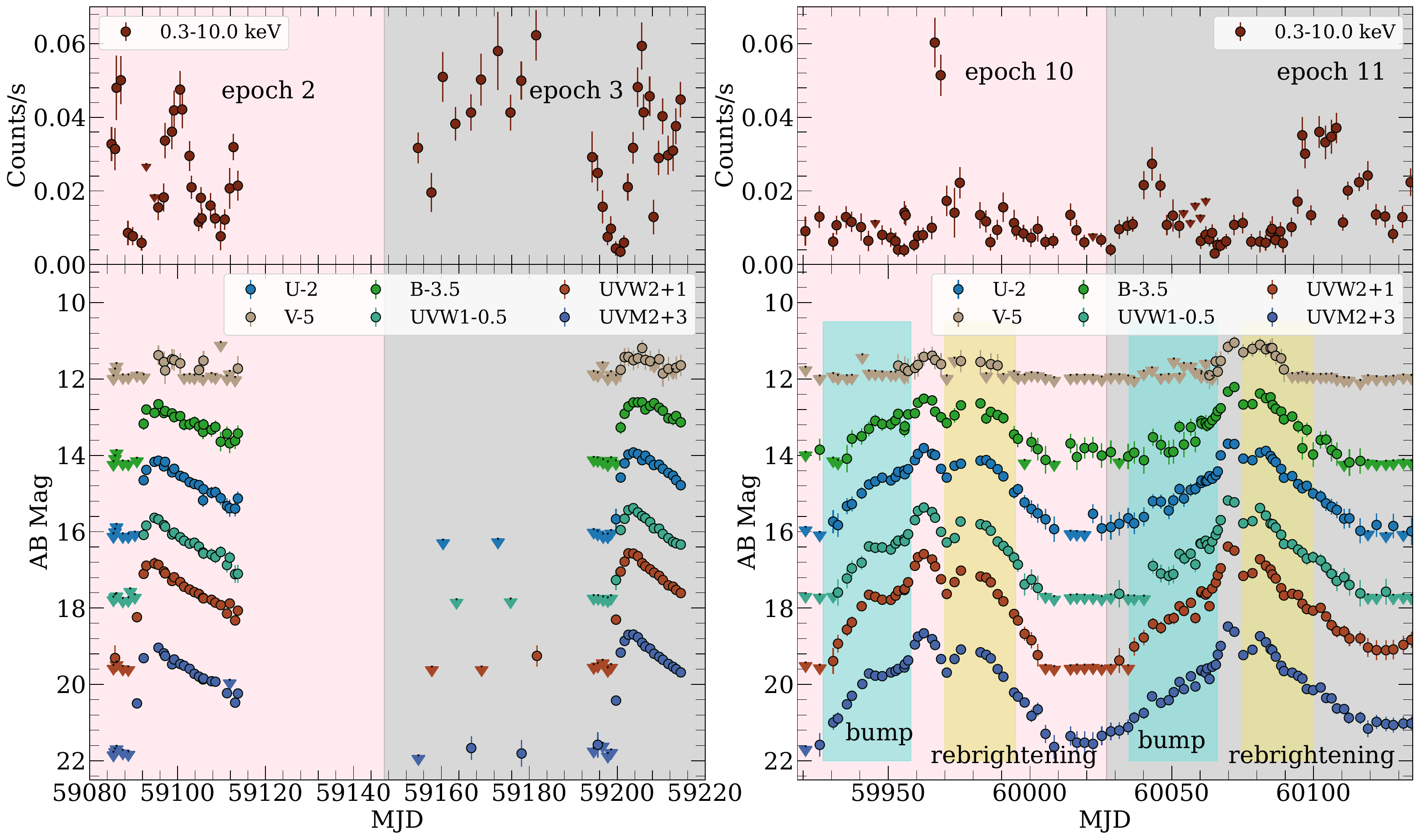}}
  \caption{Multiwavelength light curves of ASASSN-14ko. In panel (a), the historical X-ray light curve, hardness ratio, and UV/optical light curves are shown. In panel (b), we plot the details of the multiwavelength light curves of the new observations. For comparison, the light curves of epochs 2 and 3 are also plotted. The early bumps and rebrightenings are marked by a color shadow region. The magnitudes are host-subtracted. The triangles represent the 3$\sigma$ upper limit.}\label{fig:lc_mw}
\end{figure*}

\subsection{Multiwavelength light curves} The UV/optical light curves of ASASSN-14ko show a ``fast rise and slow decay'' pattern, which has been reported in previous work \citep{Payne2021,Payne2022,Payne2023}. Long-term multiwavelength light curves are shown in Figure \ref{fig:lc_mw}. For convenience, we have divided the period of MJD 58927–60110 into different epochs. As we can see, the UV/optical light curves in epochs 10 and 11 do not follow a trend similar to that in previous epochs. Interestingly, early bumps can be seen in the rising part of the UV/optical light curves in epochs 10 and 11. Additionally, the rebrightening feature is also detected in both epochs 10 and 11.  We carefully examined the light curves of previous epochs and read the detailed reports of \cite{Payne2021,Payne2022,Payne2023}, and discovered that the bump during the rising phase and the rebrightening during the declining phase is an unprecedented occurrence.
Moreover, the X-ray light curve in the interval covering epochs 10 and 11 shows signs of quasiperiodicity and we derive a probable period of 2 months using the Lomb-Scargle package \citep{VanderPlas2018}. However, more high-cadence X-ray observations are needed to confirm the periodicity.

\subsubsection{UV/Optical light curves in epoch 10} For epoch 10, the UV/optical bands start to rise from MJD 59935 and reach the early bump at MJD 59943. At the bump, the host-subtracted magnitudes are 16.12 (upper limit), 16.81, 16.77, 16.89, 16.72, 16.65 for \textsl{V}, \textsl{B}, \textsl{U}, \textsl{UVW1}, \textsl{UVM2}, and \textsl{UVW2}, respectively. The bump lasts until MJD 59952 and then the UV/optical bands continue to rise and reach the peak at MJD 59962, with the magnitudes of 16.42, 16.02, 15.81, 15.87, 15.66, 15.59 for \textsl{V}, \textsl{B}, \textsl{U}, \textsl{UVW1}, \textsl{UVM2}, and \textsl{UVW2}, respectively. After the peak, the UV/optical light curves decline until MJD 59970, with magnitudes of 16.91 (upper limit), 16.65, 16.58, 16.78, 16.69, 16.63 in \textsl{V}, \textsl{B}, \textsl{U}, \textsl{UVW1}, \textsl{UVM2}, and \textsl{UVW2}, respectively. Subsequently, a rebrightening trend is observed and the brightness increases to magnitudes of 16.53, 16.19, 16.22, 16.23, 16.09, and 16.02 (\textsl{V}, \textsl{B}, \textsl{U}, \textsl{UVW1}, \textsl{UVM2}, and \textsl{UVW2}) at MJD 59975. Unfortunately, \emph{Swift} missed the peak of the rebrightening phase.

\subsubsection{UV/Optical light curves in epoch 11}
In epoch 11, the UV/optical bands rise slowly from MJD 60034. Instead of a smooth evolution, fluctuations can be seen in the rising phase. We use the $\chi^2$ test to assess the significance of the ``fluctuations” in different bands during the rising phase. We find that the ``fluctuations”  are significant, although the data have relatively large errors. The same test is applied to the data set from epoch 10, and no significant fluctuations are found. Until MJD 60060, the UV/optical reach a short plateau with magnitudes of 16.86 (upper limit), 16.62, 16.72, 16.82, 16.63, and 16.59 in \textsl{V}, \textsl{B}, \textsl{U}, \textsl{UVW1}, \textsl{UVM2}, and \textsl{UVW2}, respectively. At MJD 60065, they start to rise and reach the UV peak at MJD 60070, with magnitudes of 16.15, 15.82, 15.70, 15.68, 15.49, and 15.39 for \textsl{V}, \textsl{B}, \textsl{U}, \textsl{UVW1}, \textsl{UVM2}, and \textsl{UVW2}, respectively. After the peak, the brightness declines and a rebrightening phenomenon is seen until MJD 60075. The peak of rebrightening is around MJD 60081, with magnitudes of 16.12, 15.90, 15.92, 15.89, 15.74, and 15.72 for \textsl{V}, \textsl{B}, \textsl{U}, \textsl{UVW1}, \textsl{UVM2}, and \textsl{UVW2}, respectively. As the light curves drop, a small bump is seen in the range of MJD 60089–60096. During the peak of this small bump, the magnitudes of \textsl{V}, \textsl{B}, \textsl{U}, \textsl{UVW1}, \textsl{UVM2}, and \textsl{UVW2} are 16.85 (upper limit), 16.51, 16.51, 16.84, 16.69, and 16.64, respectively.

\section{Analysis}\label{sec:ananlysis}
\subsection{UV/Optical Evolution in Epochs 10 and 11}
\begin{figure*}[htb]
  \centering
  \subfigure[]{
  \includegraphics[width=0.45\textwidth]{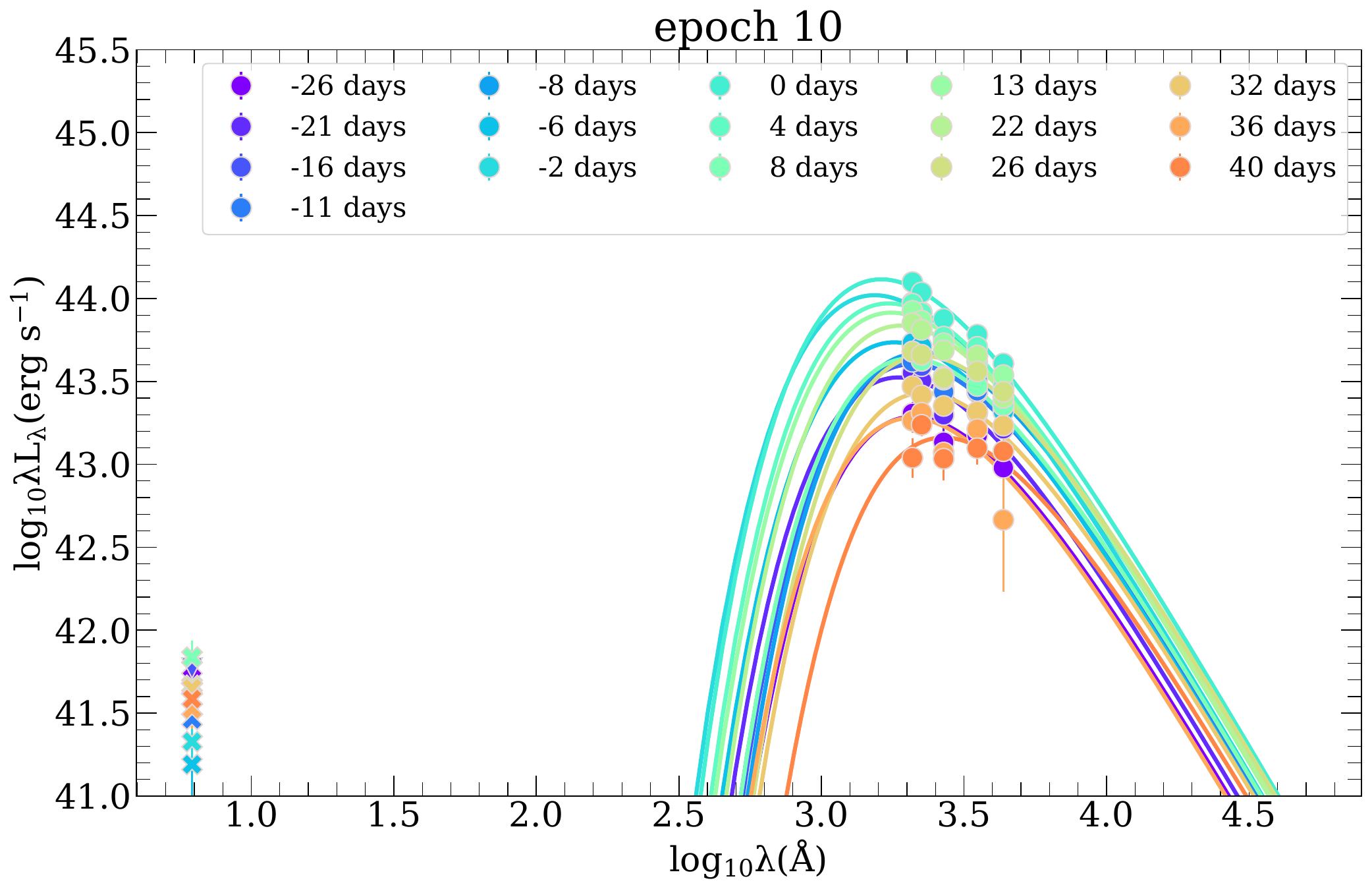}}
  \subfigure[]{
  \includegraphics[width=0.45\textwidth]{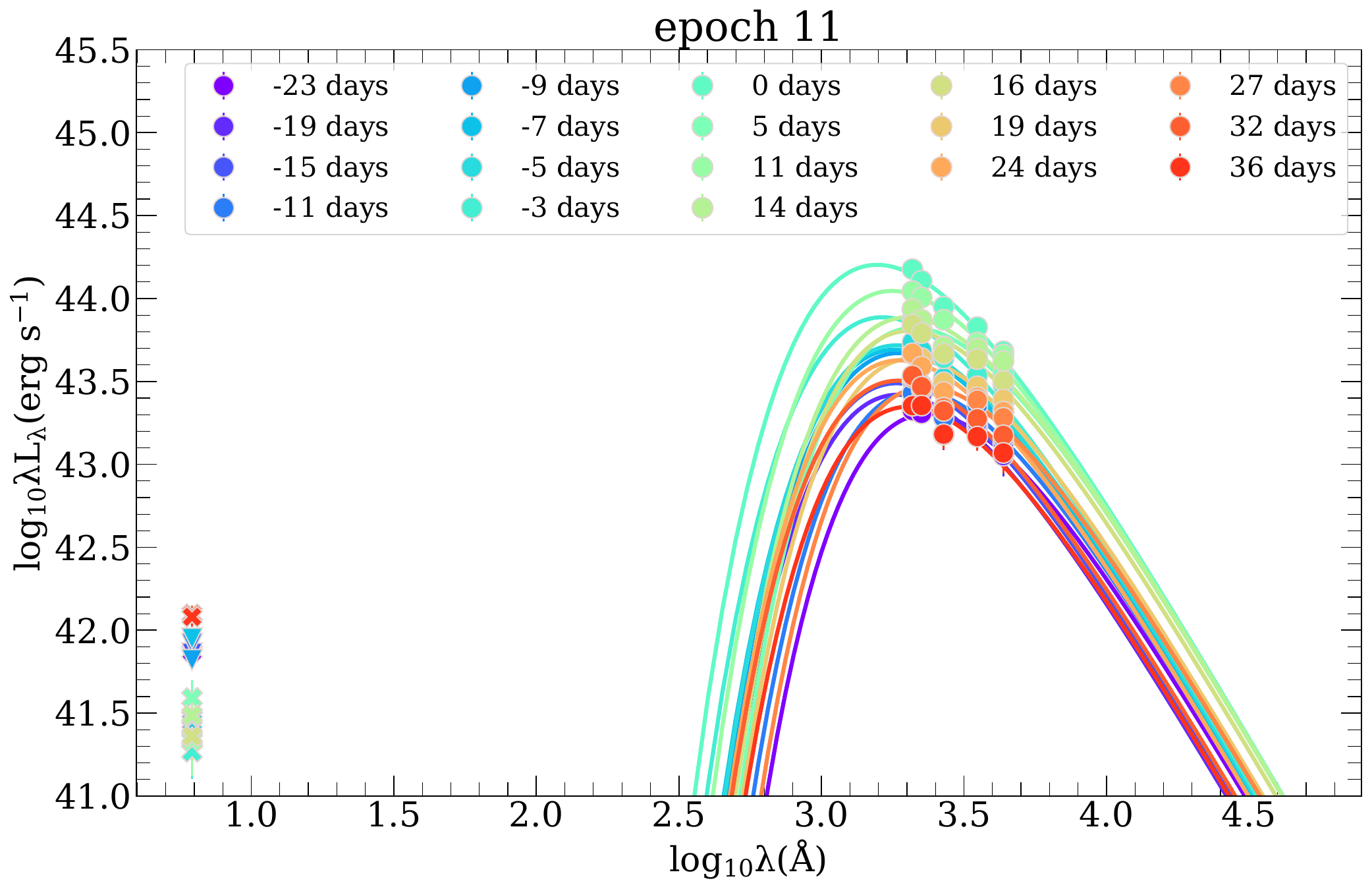}}
  \caption{SED fitting results of epochs 10 and 11. The time in the legend is relative to the UV peak. The X-ray data points at 2 keV are plotted, but are not included in the fitting. }\label{fig:sed}
\end{figure*}

\begin{figure}[htb]
  \centering
  \includegraphics[width=0.45\textwidth]{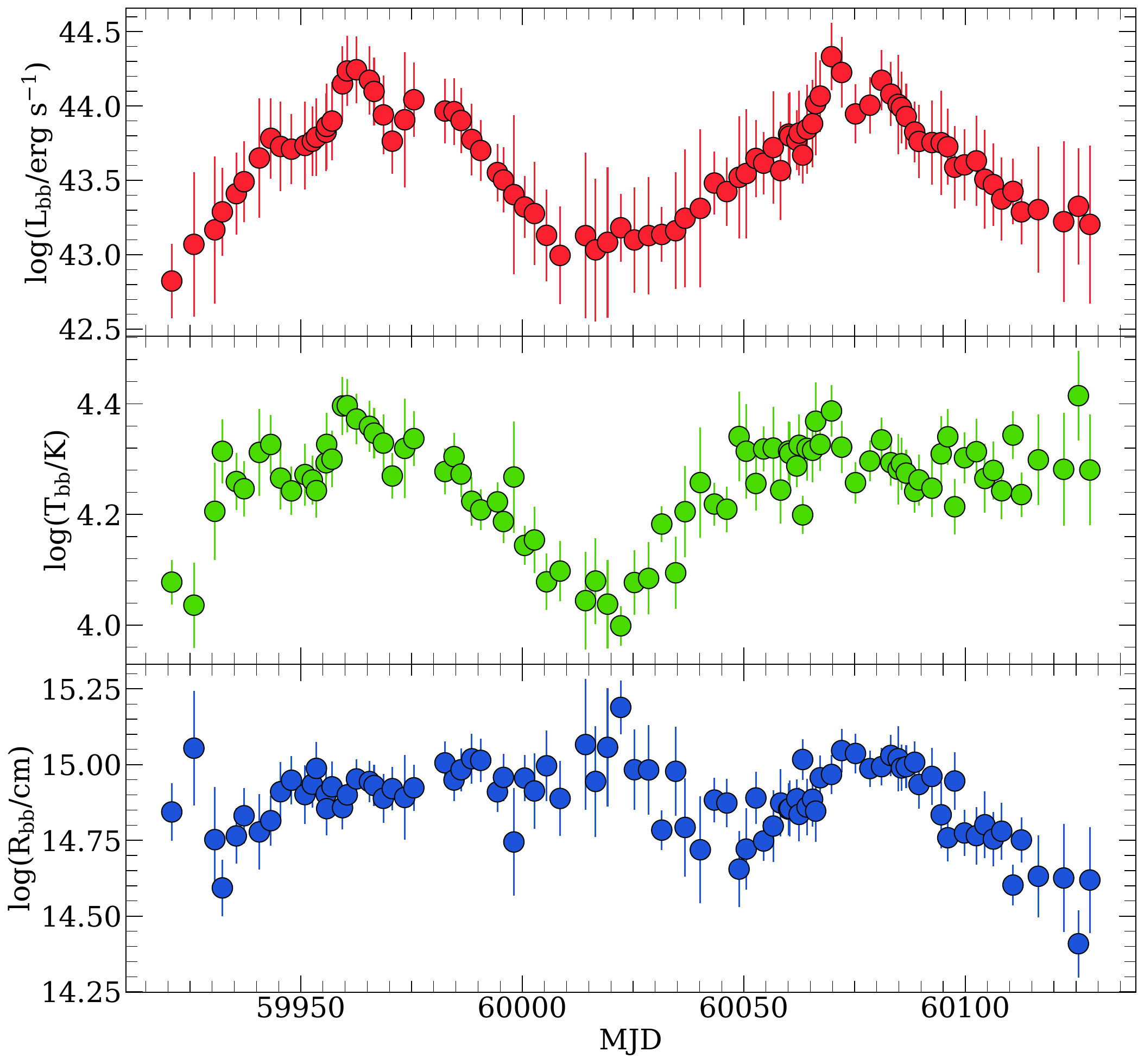}
  \caption{Evolution of UV/optical bands. Here, to reduce the impact of the large data error in the v band, we excluded this band and fitted the remaining data to derive the results. The top panel plots the evolution of host-subtracted UV/optical bolometric luminosity. The evolution of the blackbody temperature and radius are shown on the middle and bottom panels. }\label{fig:disk_temper}
\end{figure}

We use the Python package \texttt{lmfit} \citep{Newville2016} to fit the spectral energy distributions (SEDs) of the simultaneous UV/optical bands with a blackbody model, and obtain the evolution of the blackbody temperature ($\rm T_{bb}$) and radius ($\rm R_{bb}$). SED fittings are shown in Figure \ref{fig:sed}. During epochs 10 and 11, the blackbody temperature evolves with luminosity ($\rm L_{bb}$), but remains high ($>10^4$\,K) even when the source is quiescent  (around $\log(\rm L_{bb}/erg~s^{-1})=43$, within an amplitude of 0.2). The results are shown in Figure \ref{fig:disk_temper}.

In the first peak during epoch 10 (MJD 59962), we derive a peak luminosity of $\log\,(\rm L_{bb}/\rm \lum)=44.24\pm{0.22}$, with $\log\,(\rm T_{bb}/\rm K)=4.37\pm{0.05}$ and $\log\,(\rm R_{bb}/\rm cm)=14.95\pm{0.06}$. For the early bump, we derive a maximum luminosity of $\log\,(\rm L_{bb}/\rm \lum)=43.78\pm{0.27}$ in MJD 59943, with $\log\,(\rm T_{bb}/\rm K)=4.32\pm{0.05}$ and $\log\,(\rm R_{bb}/\rm cm)=14.81\pm{0.08}$. During the rebrightening phase, we derive a maximum luminosity of $\log\,(\rm L_{bb}/\rm \lum)=44.04\pm{0.25}$ in MJD 59975, with $\log\,(\rm T_{bb}/\rm K)=4.34\pm{0.05}$ and $\log\,(\rm R_{bb}/\rm cm)=14.92\pm{0.08}$.

During epoch 11, we obtain a peak luminosity of  $\log\,(\rm L_{bb}/\rm \lum)=44.33\pm{0.23}$ in MJD 60070, and the temperature and radius are $\log\,(\rm T_{bb}/\rm K)=4.39\pm{0.05}$ and $\log\,(\rm R_{bb}/\rm cm)=14.97\pm{0.06}$, respectively. On the short plateau, we derive $\log\,(\rm L_{bb}/\rm \lum)=43.81\pm{0.27}$, $\log\,(\rm T_{bb}/\rm K)=4.31\pm{0.05}$ and $\log\,(\rm R_{bb}/\rm cm)=14.85\pm{0.09}$ in MJD 60060. In the rebrightening phase of epoch 11, we derive a maximum luminosity of $\log\,(\rm L_{bb}/\rm \lum)=44.17\pm{0.20}$ in MJD 60081, with $\log\,(\rm T_{bb}/\rm K)=4.33\pm{0.04}$ and $\log\,(\rm R_{bb}/\rm cm)=14.99\pm{0.06}$.

\subsection{X-Ray behavior in Epochs 10 and 11}
During epochs 10 and 11, \emph{Swift} observes a relatively low brightness with a mean count rate of 0.01, compared to the previous epochs, which have an average count rate of 0.03. The average hardness ratio (HR)\footnote{The hardness ratio is defined by $\frac{H-S}{H+S}$, where H and S denote the count rate of the hard X-ray (2.0--10.0 keV) and the soft X-ray (0.3--2.0 keV), respectively.}  in epochs 10--11 is -0.08, while for previous epochs, the value is -0.24.

\begin{figure}[htb]
  \centering
  \includegraphics[width=0.45\textwidth]{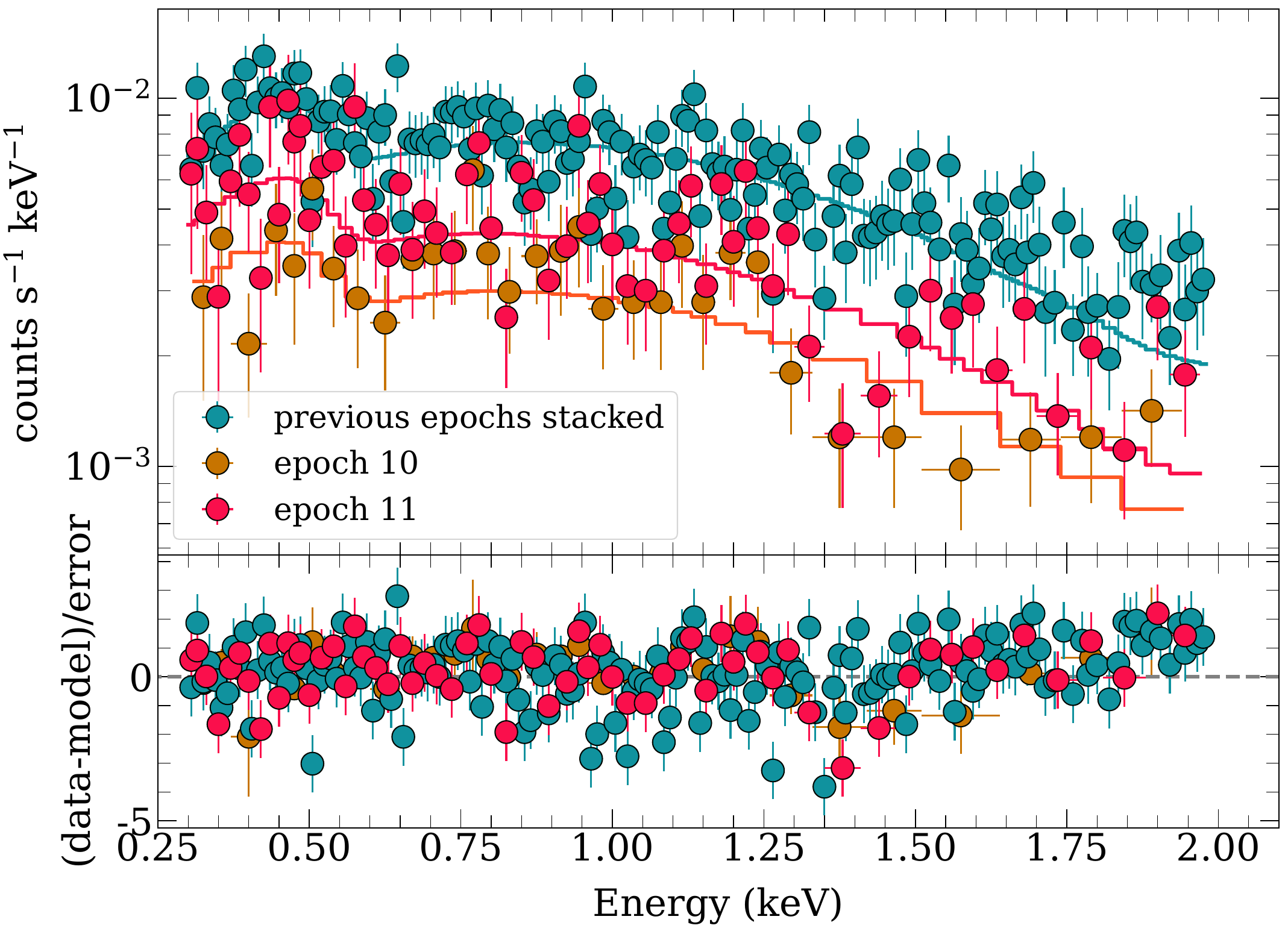}
  \caption{0.3--2.0 keV X-ray spectra in different epochs of ASASSN-14ko. All three spectra are fitted with the absorb power-law model \texttt{tbabs*zashift*powerlw}.}\label{fig:xspec}
\end{figure}

To analyze the X-ray spectra in epochs 10 and 11, we stack the spectra in the three epochs separately and group them with a minimum of 10 photons. The effective exposure times are 363, 82.5, 90.58 ks for the spectrum of previous epochs, epoch 10 and epoch 11, respectively. Through \texttt{xspec v12.13.0c}, three spectra including previous epochs, epochs 10 and 11 are fitted by an absorb power-law model (\texttt{tbabs*zashift*powerlw}). The XRT spectra exhibit the Fe K$\alpha$ line; however, \cite{Payne2023} find that the feature originates from the nearby source. Therefore, we only used the 0.3--2.0 keV range in our spectral fitting to exclude the contribution of the nearby source. Fixing a Galactic hydrogen density of $3.49\times 10^{20}\,\text{cm}^{-2}$ \citep{HI4PI2016}, we derive the photon index of $2.17\pm{0.07}$ ($\chi^2$/dof=1.42), $2.27^{+0.20}_{-0.21}$ ($\chi^2$/dof=0.92), and $2.33_{-0.16}^{+0.15}$ ($\chi^2$/dof=1.20) for the results of 0.3--2.0 keV in previous epochs, epoch 10 and epoch 11, respectively. The spectra in epochs 10 and 11 are slightly softer than those in the previous epochs, but no significant difference in profile is found in these X-ray spectra. Due to the inability of \emph{Swfit}/XRT to distinguish the source adjacent to ASASSN-14ko in the image, we limit ourselves to a qualitative comparison here. The fitted results are shown in Figure \ref{fig:xspec}.

\begin{figure*}[htb]
  \centering
  \subfigure[]{
  \includegraphics[width=0.45\textwidth]{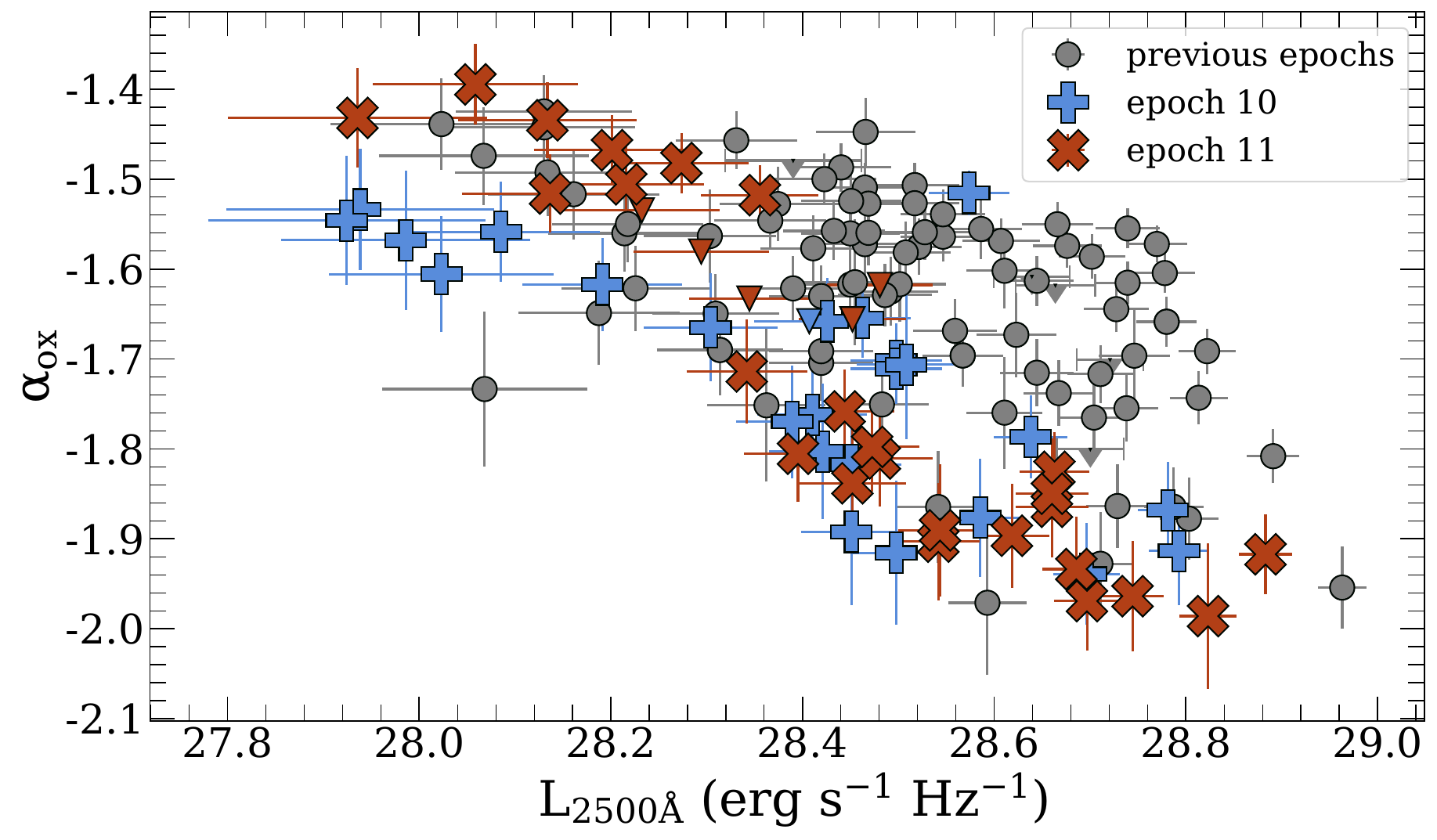}}
  \subfigure[]{\includegraphics[width=0.45\textwidth]{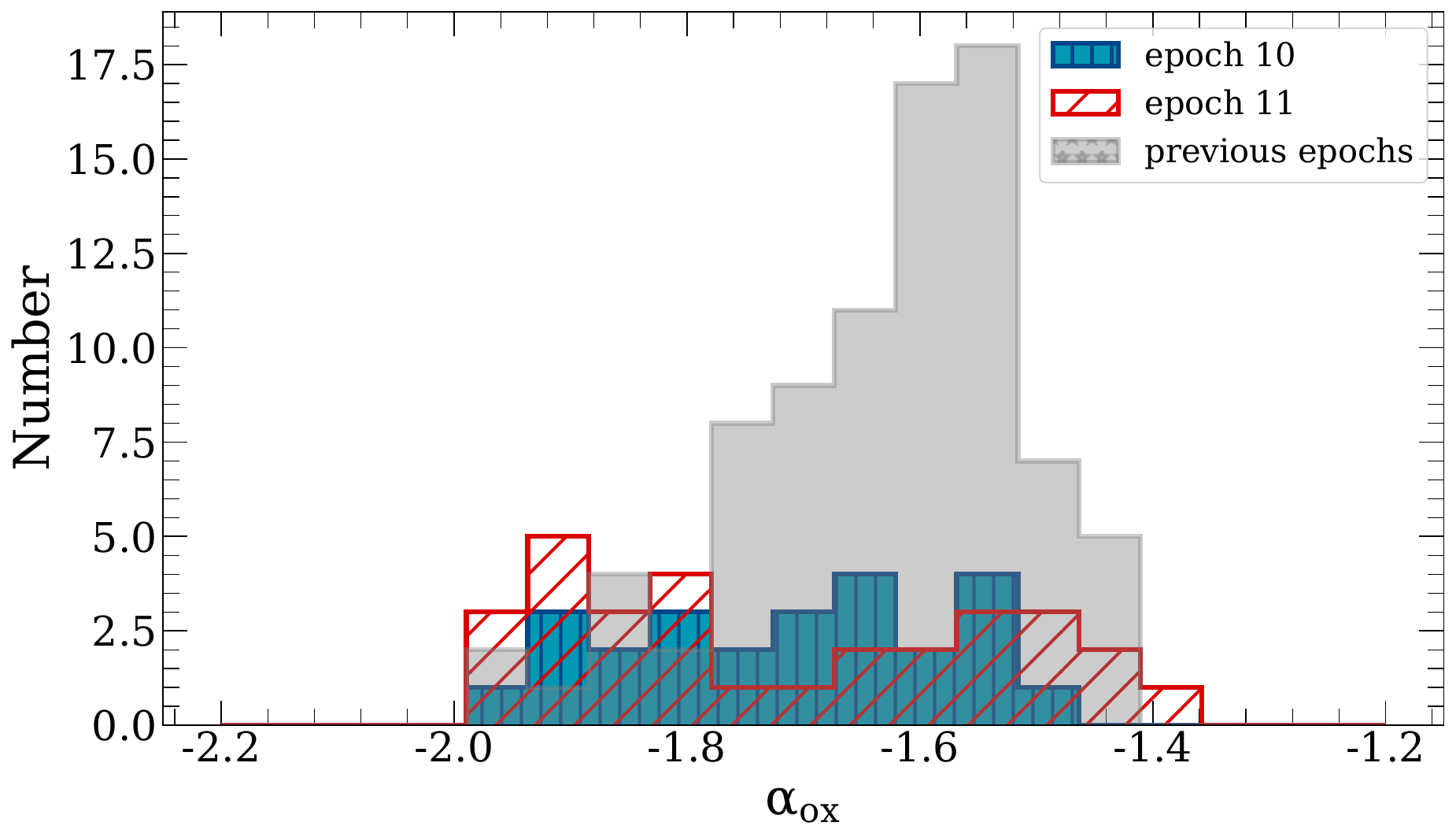}}
  \caption{(a), the correlation between $\alpha_{\text{ox}}$ and the luminosity of {2500 \AA} in different epochs. (b), the distribution of $\alpha_{\text{ox}}$ at different epochs}\label{fig:aox}
\end{figure*}

It should be noted that the variability of the X-rays is not consistent with the UV/optical bands. The X-ray and UV/optical fluxes show anticorrelated behavior in epochs 10 and 11, as they did in previous epochs. When the UV/optical fluxes increase, the X-ray fluxes decrease, and vice versa. This phenomenon has been observed by \cite{Payne2021,Payne2022,Payne2023} Therefore, we suggest that the X-ray flare near the UV/optical bumps and rebrightenings is not a new feature, but a continuation of the previous epoch. To measure the X-ray to UV/optical flux ratio, we use the parameter $\alpha_{\text{ox}}=0.3838~\log\left[\frac{\text{F}(\text{2 keV})}{\text{F(2500\AA)}}\right]$, where F(\text{2 keV}) and F(2500\AA) are the flux at 2 keV and 2500\AA, respectively \citep{Tananbaum1979,Strotjohann2016}. The values of $\alpha_{\rm ox}$ range from -1.3 to -2.  In Figure \ref{fig:aox},  $\alpha_{\text{ox}}$ in epochs 10 and 11 lie mainly on different sides from previous epochs. For previous epochs, $\alpha_{\rm ox}$ mostly ranges from -1.8 to -1.4, while for epochs 10 and 11, it is lower. Especially for epoch 11, we find a dominant distribution of $\alpha_{\rm ox}$ values between -2.0 and -1.6.  For $\log(\rm L_\text{2500\AA}/\rm erg\,\rm s^{-1}\,\rm Hz^{-1}) > 28.3$, $\alpha_{\rm ox}$ in epochs 10 and 11 tends to be lower, where $\rm L_\text{2500\AA}$ is the luminosity in {2500 \AA}. The result suggests a distinct correlation between the X-ray and UV/optical bands in epochs 10 and 11.

\subsection{Energy release in UV/Optical bands}  \label{sec:energy}
Since the light curves in the previous epochs showed a smooth ``fast-rise and slow-decay" pattern, we assume that they resulted from the same mechanism and that the folded light curves in the previous epochs can serve as a baseline for the analysis. Therefore, we can derive the extra energy in epochs 10 and 11 by subtracting the energy in the previous epochs.

The total energy released in epoch 10 is $4.74\times10^{50}\,\rm erg$, and in epoch 11, the value of $4.66\times10^{50}\,\rm erg$ is reached, while the average energy release in the previous epochs is $2.64\times10^{50}\,\rm erg$. Compared to previous epochs, the extra energy released in the early bumps of epochs 10 and 11 is $1.15\times 10^{50}\,\rm erg$ and $9.15\times 10^{49}\,\rm erg$, respectively. For the rebrightening phases, the extra energy released in epochs 10 and 11 is $7.63\times 10^{49}\,\rm erg$ and $7.57\times 10^{49}\,\rm erg$, respectively. The energy released in epoch 10 slightly exceeds that in epoch 11, but both are significantly higher than the average values of previous epochs.

\section{Theoretical interpretation}\label{sec:diss}

The UV/optical light curves of the newly observed epochs 10 and 11 appear to deviate from the pattern observed in previous epochs, \footnote{Because the X-ray luminosity is lower than UV/optical by one order of magnitude, here we mainly focus on the latter.} indicating anomalies in the behavior of ASASSN-14ko. Unlike the previous epochs, which exhibited a consistent "fast rise and slow decay" pattern in their UV/optical light curves during each outburst~\citep{Payne2021,Payne2022,Payne2023}, epochs 10 and 11 show a much slower rise, taking more than 1 month to reach their peaks. This longer rise can be attributed to the presence of a bump structure, which was not observed in the earlier epochs. A similar bump structure was also observed in the rising phase of AT2020wey, although it was fainter compared to ASASSN-14ko~\citep{Charalampopoulos2023}. Another noteworthy phenomenon is the rebrightenings in the declining UV/optical light curves of epochs 10 and 11. While rebrightenings in TDEs are common, these recent rebrightenings in ASASSN-14ko are particularly intriguing as they occur in a periodic source that previously exhibited a rather smooth decline in its outbursts. Therefore, the rebrightening phenomenon observed in ASASSN-14ko might have a different origin compared to rebrightenings reported in other TDEs.


\subsection{Stripped Stream-Disk Collision in the Partial TDE Scenario}

The partial TDE model is thus far the primary model adopted to explain the periodic outbursts of ASASSN-14ko~\citep{Payne2021,Liu2023}. According to this model, a star in an eccentric orbit around the black hole undergoes tidal stripping each time at the pericenter $R_{\rm p}$. The stripped material then undergoes circularization and subsequent accretion, resulting in the observed flare.

Tidal stripping requires $R_{\rm p}$ to be on the same order as the tidal radius, i.e., $R_{\rm p}= \chi R_* (M_h/M_*)^{1/3}$, where $\chi \ge 1$ is a numerical coefficient; this is the distance to the SMBH at which a tiny amount of mass could be removed from the star. At this point, the star just fills its Roche lobe (though instantaneously due to the eccentric nature of the orbit). Thus, $\chi \simeq 2$ according to early work \citep{Paczynski1971,Eggleton1983,Sepinsky2007}. Recent numerical simulation and analytical studies of TDE suggest similar values: $\chi= 1.7 \sim 2$ \citep{Guillochon2013,Coughlin2022}. Note that a complete disruption would actually correspond to $\chi \le 1$). Here, we adopt $\chi_s=2$. Thus, $R_{\rm p}=2.2 \, M_8^{-2/3}r_{*}m_{*}^{-1/3}R_{\rm S}$, where $M_8$ is the mass of the SMBH in unit of $10^8M_\sun$, $r_{*}$ is the stellar radius in unit of solar radius, $m_{*}$ is the stellar mass in unit of solar mass and $R_{\rm S}$ is the Schwarzschild radius of the SMBH. Note that \cite{Payne2021} derived the SMBH mass of $10^{7.85}\, M_\odot$ for ASSASN-14ko. The semi-major axis of the stellar orbit is related to the orbital period as in $a= 110\, (P/114\, \mbox{d})^{2/3} M_8^{-2/3} \,R_{\rm S}$.

\begin{figure*}[htb]
  \centering
  \includegraphics[width=0.7\textwidth]{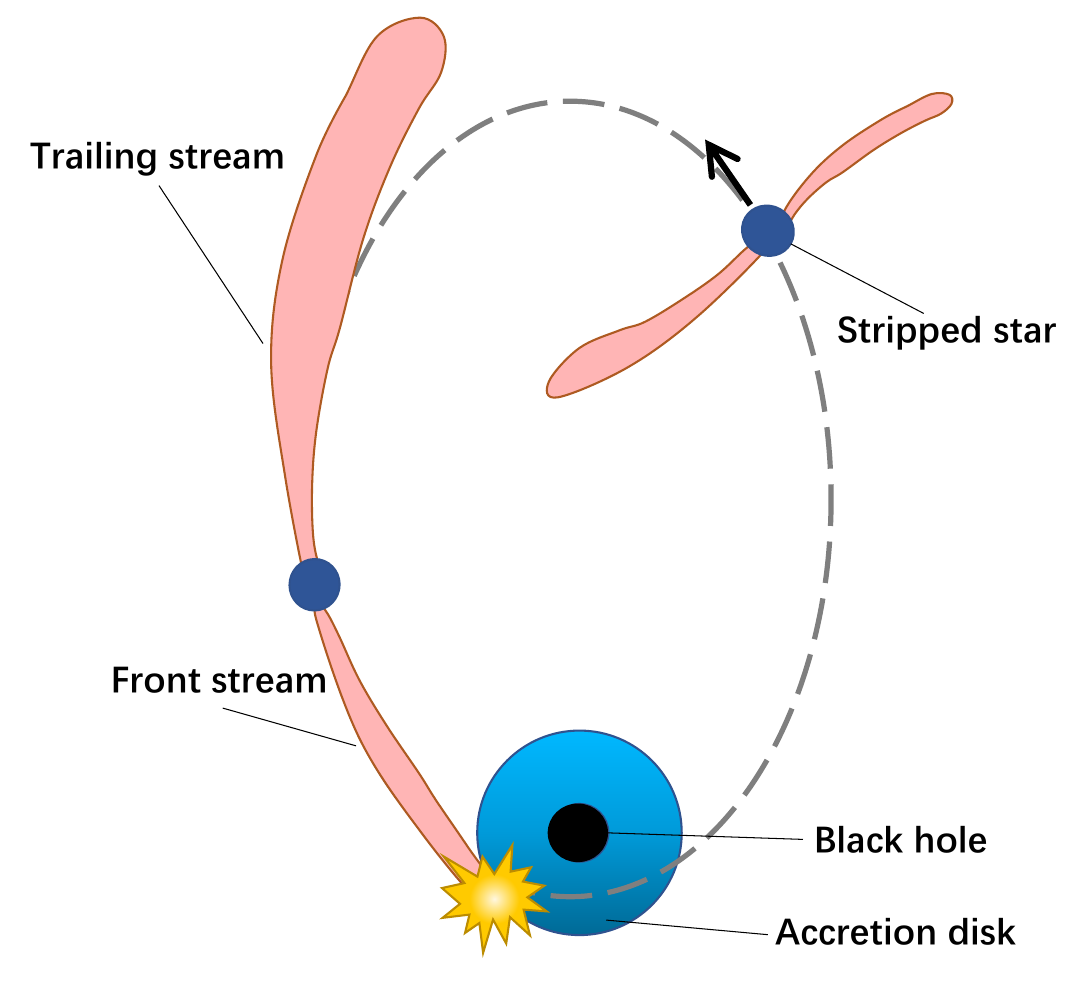}
  \caption{The star is tidally stripped by the black hole, creating a debris stream that orbits around it. The stream crashes into the accretion disk twice in every cycle. These collisions produce flares in the UV/optical bands. The first flare occurs when the front stream hits the disk, creating an early bump in the light curves. The second flare occurs when the trailing stream collides with the disk, causing a rebrightening of the emission. The main outburst is powered by the accretion of the stripped material onto the black hole.
}\label{fig:model}
\end{figure*}

The unexpected profile change of the light curves during the most recent outbursts poses a challenge to the partial TDE explanation. Why were the bump in the rising phase and the rebrightening in the declining phase absent in the past?
We propose the hypothesis that an accumulation effect is responsible for the two repeated bumps and rebrightenings. In this hypothesis, not all debris fragments from the partial TDE are immediately accreted, but instead some remain and gradually accumulate over time. As these materials accumulate from multiple partial TDEs, the accretion disk grows larger. Consequently, a collision occurs between the stripped stream and the disk, resulting in the early bumps and rebrightenings both before and after the flare peak.  The schematic representation of the model can be seen in Figure \ref{fig:model}. We will quantitatively examine and evaluate this model in the subsequent analysis.

At the collision radius $R_{\rm p}$, for a disk thickness $H \gg R_*$, the length of the penetration of the stream through the disk would be $l \sim R_{\rm p}$. The stream-disk velocity difference is $v_{\rm rel} \sim \sqrt{2 G M_h /R_{\rm p}}$. The disk material to be shocked is in a volume of $\pi R_*^2 R_{\rm p}$, whose mass is $\Delta M_{\rm s} = \pi R_*^2 R_{\rm p} \rho$
where $\rho$ is the vertically averaged density of the disk, which is related to the disk accretion rate as $\dot{M}_{\rm acc}= 3 \pi \nu \rho H$. Adopting the Shakura -- Sunyaev viscosity model $\nu= 2 \alpha P /(3 \Omega_{\rm K}  \rho)$ and utilizing $H= (P/\rho)^{1/2}/\Omega_{\rm K}$, it can be rewritten as
\begin{equation}   \label{eq:mdotacc}
\dot{M}_{\rm acc}= 2\pi R^2 \rho H \alpha \Omega_{\rm K} \left(\frac{H}{R}\right)^2,
\end{equation}
where $\Omega_{\rm K} (R)$ is the disk Keplerian angular speed. With the density from Eq. (\ref{eq:mdotacc}), we get the mass of the shocked material $\Delta M_{\rm s}$.
An optimistic estimate of the energy dissipated would be $\Delta E_{\rm s} \approx \Delta M_{\rm s} v_{\rm rel}^2/2$.
Then we get
\begin{equation}
\begin{split}
\Delta E_{\rm s}= & 8\times10^{49} \, r_*^{1/2} \left(\frac{\dot{m}_{\rm acc}}{0.1} \right)  \left(\frac{\alpha}{0.1}\right)^{-1} \\ & \times \left(\frac{H/R}{0.01}\right)^{-3} M_8 m_*^{1/2} \, \mbox{erg},
\end{split}
\end{equation}
where the accretion rate has been normalized by the Eddington rate as in $\dot{m}_{\rm acc} \equiv 10 \, \dot{M}_{\rm acc} c^2/ L_{\rm Edd}$. To match the observed energies for the early bumps (see \ref{sec:energy}), we chose a low (but not extreme) value for the disk height-to-radius ratio.

We note that the separation between the bump and the rebrightening in epoch 11 ($\Delta \rm t_{11}$) is shorter than that in epoch 10 ($\Delta \rm t_{10}$). The orbital precession of the stripped stream may cause the inequality between $\Delta \rm t_{10}$ and $\Delta \rm t_{11}$. The stream impacts the accretion disk twice in each period but at different positions due to the precession. This leads to unequal intervals between the two impacts in each period. This explains why we observe $\Delta \rm t_{10}\neq\Delta \rm t_{11}$. A similar situation occurs in OJ287, except that in that source it is the secondary black hole that collides with the accretion disk \citep{Valtonen2008,Valtonen2016,Dey2018,Laine2020}.

For partial TDE, the outer radius of the disk is $R_{\rm out} \simeq 2 R_{\rm p}$, and in the case of ASASSN-14ko, $R_{\rm out}=5.5R_{\rm S}$. Such a short distance to the SMBH makes it difficult to accumulate the stream debris for a long time, and the expansion of the disk is improbable. However, for an evolved star whose envelope is extended $R_* \sim 10R_\sun$, as advocated by \cite{Liu2023} for ASASSN-14ko, the disk size is much larger $R_{\rm out}=55\, R_{\rm S}$. At this distance, the accretion disk has the possibility of expanding. Furthermore, a larger $R_*$ also helps increase the energy release for the collision.

\subsection{Other Scenarios}

We explore alternative explanations that could account for the distinct periodic transient ASASSN-14ko, as well as the newly observed features. An abnormal behavior for the TDE scenario is the significant temperature evolution during the outburst (see Figure~\ref{fig:disk_temper}), in contrast to the relatively stable temperature observed in normal optical TDEs. This temperature fluctuation bears resemblance to the disk instability scenario. Particularly, we noticed that the disk instability model has been proposed to explain QPEs in previous studies \citep{Pan2023, Sniegowska2023} and the X-ray flares observed in ASASSN-14ko are rather similar to QPEs. Another popular model involved in explaining QPE is the star-disk collision model~\citep{Xian2021,Linial2023}, which is yet disfavored by a lower energy released by 1 order of magnitude than that observed in individual flares of ASASSN-14ko ($\sim 10^{50}$ erg). One potential explanation is that the star periodically intersects with the disk, triggering disk instability and giving rise to the prominent flares observed.  In this case, the star has just been disrupted; the stripped material, along with the star itself, falls back onto the disk in two streams. These streams then collide with the accretion disk, resulting in the observed bumps before and after the main flare peaks.

\section{Conclusion}
ASASSN-14ko is a rare event that displays periodic outbursts in UV/optical bands, which are likely caused by repeated partial TDEs. The two most recent observational epochs, reported in this paper, reveal that the last two flares show much extended temporal profiles, characterized by early bumps and postpeak rebrightenings. This new feature could potentially be explained by the additional energy dissipation from the plunging of the tidally stripped stream through an expanded accretion disk. The periodic nature of ASASSN-14ko provides us a unique and valuable opportunity to study the physical processes underlying TDEs by conducting well-designed multiwavelength observations, both in the most recent epochs and in the near future.

\begin{acknowledgments}
 We thank the anonymous referee for providing valuable comments, which help to improve the quality of this work.
This work is supported by the SKA Fast Radio Burst and High Energy Transients Project (2022SKA0130102), the National Natural Science Foundation of China (grants 11833007, 12073025, 12192221, 12103048, 12233008, 12073091), Anhui Provincial Natural Science Foundation (2308085QA32) and the Fundamental Research Funds for Central Universities (WK3440000006). We gratefully acknowledge the support of Cyrus Chun Ying Tang Foundations, the Frontier Scientific Research Program of the Deep Space Exploration Laboratory (2022-QYKYJH-HXYF-012).
The authors acknowledge the use of public data from the Swift data archive. The authors thank the Swift ToO team for accepting our proposal and executing the observations.
\end{acknowledgments}

\vspace{5mm}
\facilities{Swift/XRT and Swift/UVOT}

\software{astropy \citep{2013A&A...558A..33A}, HEASoft \citep{2014ascl.soft08004N}, Xspec \citep{1996ASPC..101...17A}, Matplotlib \citep{2007CSE.....9...90H}, lmfit \citep{Newville2016}.
          }

\bibliography{ASASSN-14ko}{}
\bibliographystyle{aasjournal}

\end{document}